\documentclass[twocolumn,showpacs,preprintnumbers,superscriptaddress]{revtex4}

\usepackage{amsfonts,amssymb,amsmath,amsthm}

\advance\topmargin 1.4cm

\newtheorem{theorem}{Theorem}
\newcommand{\calA}{{\cal A }}
\newcommand{\calZ}{{\cal Z }}
\newcommand{\calC}{{\cal C }}
\newcommand{\la}{\langle}
\newcommand{\ra}{\rangle}
\newcommand{\be}{\begin{equation}}
\newcommand{\ee}{\end{equation}}

\newcommand{\ba}{\begin{array}}
\newcommand{\ea}{\end{array}}

\newcommand{\rhosub}[1]{\rho_{\scriptscriptstyle #1}}
\newcommand{\rhosubsup}[2]{\rho_{\scriptscriptstyle #1}^{\scriptscriptstyle #2}}
\newcommand{\psiLR}{|\psi\rangle_{LR}}

\newcommand{\Tsubsup}[2]{T_{\scriptscriptstyle #1}^{(#2)}} 

\usepackage{graphicx}
\usepackage{dcolumn}
\usepackage{bm}

\DeclareMathOperator*{\trace}{\mathop{\mathrm{Tr}}}
\DeclareMathOperator*{\nb}{\mathop{\mathrm{neigh}}}
\usepackage{color}
\usepackage{epsfig}

\newcommand{\ncd}{\newcommand}

\ncd{\QC}{$\mbox{QC}_{\cal{C}}\;$}
\ncd{\QCpr}{${\mbox{QC}_{\cal{C}}}^\prime\;$}
\ncd{\QCns}{$\mbox{QC}_{\cal{C}}$}
\ncd{\QCprns}{${\mbox{QC}_{\cal{C}}}^\prime$}
\ncd{\ds}{\displaystyle}
\ncd{\ovl}{\overline}
\ncd{\iden}{1 \hspace{-1.0mm}  {\bf l}}

\begin{document}

\title{Long-range quantum entanglement in noisy cluster states}

\author{Robert Raussendorf, Sergey Bravyi and Jim Harrington}
\affiliation{California Institute of Technology,\\
Institute for Quantum Information, Pasadena, CA 91125, USA}

\date{\today}

\begin{abstract}
  We describe a phase transition for long-range entanglement in a 
  three-dimensional cluster state affected by noise. The partially decohered 
  state is modeled by the thermal state of a short-range 
  translation-invariant Hamiltonian. We find 
  that the temperature at which the entanglement length changes from infinite 
  to finite is nonzero. We give an upper and lower bound to this transition 
  temperature.  
\end{abstract}

\pacs{3.67.Lx, 3.67.-a}

\maketitle


\section{Introduction}
\label{intro}

Nonlocality is an essential feature of quantum mechanics, put to the
test by the famous Bell inequalities \cite{Bell} and verified in a
series of experiments, see e.g. \cite{Aspect}. Entanglement
\cite{Schr} is an embodiment of this nonlocality which has become a
central notion in quantum information theory.

In realistic physical systems, decoherence represents a formidable but
surmountable obstacle to the creation of entanglement among far
distant particles. Devices such as quantum repeaters \cite{Rep} and
fault-tolerant quantum computers are being envisioned in which the
entanglement length \cite{LE,Aha} is infinite, provided the
noise is below a critical level. Here we are interested in the
question of whether an infinite entanglement length can also be found 
in spin chains with a short-range interaction that are subjected to noise.  
A prerequisite for
our investigation is the existence of systems with infinite
entanglement length at zero temperature. An example of such behavior
has been discovered by Verstraete, Mart\'in-Delgado, and
Cirac~\cite{Verstraete_Delgado_Cirac_04} with spin-1 chains in the
AKLT-model \cite{AKLT}, and by Pachos and Plenio with cluster Hamiltonians 
\cite{PP}; see also \cite{Kay}. In this paper, we study the case of finite
temperature. We present a short-range, translation-invariant Hamiltonian 
for which the entanglement
length remains infinite until a critical temperature $T_c$ is
reached. The system we consider is a thermal cluster state in three
dimensions. We show that the transition from infinite to finite
entanglement length occurs in the interval $0.30\, \Delta \leq T_c
\leq 1.15\,\Delta$, with $\Delta$ being the energy gap of the
Hamiltonian.
 
We consider a simple 3D cubic lattice ${\cal{C}}$ 
with one spin-$1/2$ particle (qubit) living 
at each vertex  of the lattice.
Let $X_u$, $Y_u$, and $Z_u$ be the Pauli operators acting 
on the spin at a vertex $u\in \calC$. The model Hamiltonian is
\begin{equation}\label{Hmltn}
H= - \frac{\Delta}{2}\, \,
\sum_{u\in {\cal{C}}}  K_u,
\quad  
 K_u = X_u \!\!\!\!\prod_{v\in \nb{(u)}}
\!\!Z_v.
\end{equation}
Here $\nb{(u)}$ is a set of nearest neighbors of 
vertex $u$. The ground state of $H$ obeys eigenvalue equations
$K_u |\phi\ra_{\calC} = |\phi\ra_{\calC}$
and coincides with a cluster state~\cite{BR}.  
We define a thermal cluster state at a temperature $T$ as
\be
\rhosub{CS} = \frac1{\calZ} \exp{( -\beta H )}, 
\ee
where $\calZ=\trace{ e^{-\beta H}}$ is a partition function and $\beta\equiv T^{-1}$.
Since all terms in $H$ 
commute, one can easily get
\begin{equation}\label{CS}
    \rhosub{CS} = \frac{1}{2^{|{\cal{C}}|}} 
      \prod_{u\in {\cal{C}}}
    \left( I + \tanh\left(\beta\Delta/2\right)\, K_u \right).
\end{equation}

Let $A, B \subset \calC$  be two distant regions on the lattice.
Our goal is to create as much entanglement between $A$ and $B$ as possible by
doing local measurements on all spins not belonging to $A\cup B$.
Denote $\alpha$ as the list of all outcomes obtained in these measurements
and  $\rhosubsup{\alpha}{AB}$ as the state of $A$ and $B$ conditioned on the
outcomes $\alpha$. Let $E[\rho]$ be some measure of bipartite entanglement.
Following~\cite{LE} we define
the localizable entanglement between $A$ and $B$ as
\begin{equation}\label{LE}
E(A,B)= \max \sum_\alpha p_\alpha\, E[ \rhosubsup{\alpha}{AB} ],
\end{equation}
where $p_\alpha$ is a probability to observe the outcome $\alpha$ and the
maximum is taken over all possible patterns  of local measurements.
To specify the entanglement measure $E[\rho]$ it is useful to regard 
$\rhosubsup{\alpha}{AB}$ as an encoded two-qubit
state with the first logical qubit residing in $A$ and the second in $B$.
We choose $E[\rho]$ as the  maximum amount of two-qubit entanglement
(as measured by entanglement of formation) contained in $\rho$. 
Thus $0\le E(A,B)\le 1$ and an equality $E(A,B)=1$ implies that
a perfect Bell pair can be created between $A$ and $B$. Conversely,
$E(A,B)=0$ implies that any choice of a measurement pattern produces a 
separable state.

In this paper we consider a finite 3D cluster 
\[
\calC = \{ u=(u_1,u_2,u_3) \, : \, 1\le u_1,u_2+1 \le l; \, \,  
1\le u_3 \le d\}\]
and choose a pair of opposite 2D faces as $A$ and $B$:
\[
A = \{ u \in \calC \, : \, u_3=1\}, \quad
B = \{ u \in \calC \, : \, u_3=d\},
\]
so that the separation between the two regions is $d-1$.
In Section~\ref{sec:LB} we show  
that~\footnote{Refs. \cite{WHP,PT_Nishimori} consider a lattice with proportions of a cube, corresponding to $l=d$. However, numerical simulations
indicate that $\lim_{l,d\to \infty} E(A,B)=1$  even if
$l=C\ln{(d)}$; see remarks to Section~\ref{sec:LB}.}
\[
\lim_{l,d\to \infty} E(A,B)=1 \quad \mbox{for} \quad  T < 0.30\, \Delta.
\]
Further, we show in Section~\ref{sec:UB} that
if $T > 1.15\,\Delta$ then 
$E(A,B)=0$ for $d\geq2$
and arbitrarily large  $l$.
 
\section{Lower bound}
\label{sec:LB}

We relate the lower bound on the transition temperature to quantum 
error correction. From Eq.~(\ref{CS}) it follows that 
$\rhosub{CS}$ can be prepared from the perfect cluster state $|\phi\ra_{\calC}$
by applying the Pauli operator $Z_u$ to each spin $u\in \calC$ 
with a  probability 
\be
\label{Eprob}
p=\frac{1}{1+\exp(\beta\Delta)}. 
\ee
Thus, thermal
fluctuations are equivalent to independent local $Z$-errors with an error rate
 $p$. 

We use a
single copy of $\rhosub{CS}$ and apply a specific pattern of local
measurements  which creates an
encoded Bell state among sets of particles in $A$ and $B$. For encoding we use 
the planar 
code, which belongs to the family of surface codes introduced by Kitaev. 
The 3D cluster state has, as opposed to its 1D counterpart \cite{BR}, an 
intrinsic error correction capability which we use in the measurement pattern described below. Therein, the measurement outcomes are individually random but not independent; parity constraints exist among them. The violation of any of these indicates an error. Given sufficiently many such constraints, the measurement outcomes specify a syndrome from which typical errors can be reliably identified. The optimal error correction given this syndrome breaks down at a certain error rate (temperature), and the Bell correlations can no longer be mediated. This temperature is a lower bound to $T_c$, because in principle there may exist a more effective measurement pattern.

To describe the measurement pattern we use, let us introduce two 
cubic sublattices $T_e, T_o\subset \calC$ with a double spacing.
Each qubit $u\in \calC$ becomes either a vertex or an edge in one of
the sublattices $T_e$ and $T_o$. The sets of vertices $V(T_e)$ and  $V(T_o)$ 
are defined as
\[
\ba{rcl}
V(T_e) &=& \{ u=(e,e,e) \in \calC \},\\
V(T_o) &=& \{ u=(o,o,o) \in \calC \},\\
\ea
\]
where  $e$ and $o$ stand for even and odd coordinates. 
The sets of edges $E(T_e)$ and $E(T_o)$ are defined as
\[
\ba{rcl}
E(T_e) &=& \{ u=(e,e,o),(e,o,e),(o,e,e) \in \calC\},\\
E(T_o) &=& \{ u=(o,o,e),(o,e,o),(e,o,o) \in \calC\}.\\
\ea
\]
The lattices $T_e$, $T_o$ play an important role in
the identification of error correction on the cluster state with a
$\mathbb{Z}_2$ gauge model \cite{TCDS}. They are displayed in 
Fig.~\ref{measpatt}

\begin{figure}
  \begin{center}
    \epsfig{file=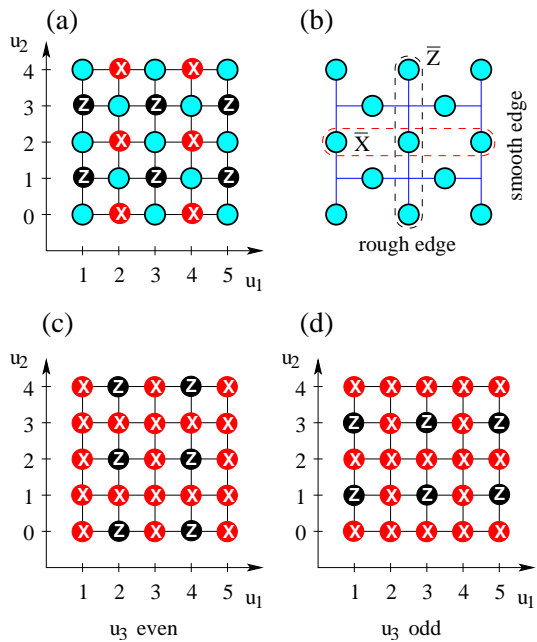, width=7cm}
    \caption{\label{OneSlice}(color online)
      (a) Measurement pattern on the first 
      and last slice of 
      ${\cal{C}}$, for $l=5$. The resulting state is in the 
    code space of the planar 
      code. (The unmeasured qubits are displayed as shaded circles.) (b) 
      Lattice for the planar code. (c)/(d) Measurement pattern for 
	even and odd inner slices.}
  \end{center}
\end{figure}

\begin{figure}[ht]
  \begin{center}
    \epsfig{file=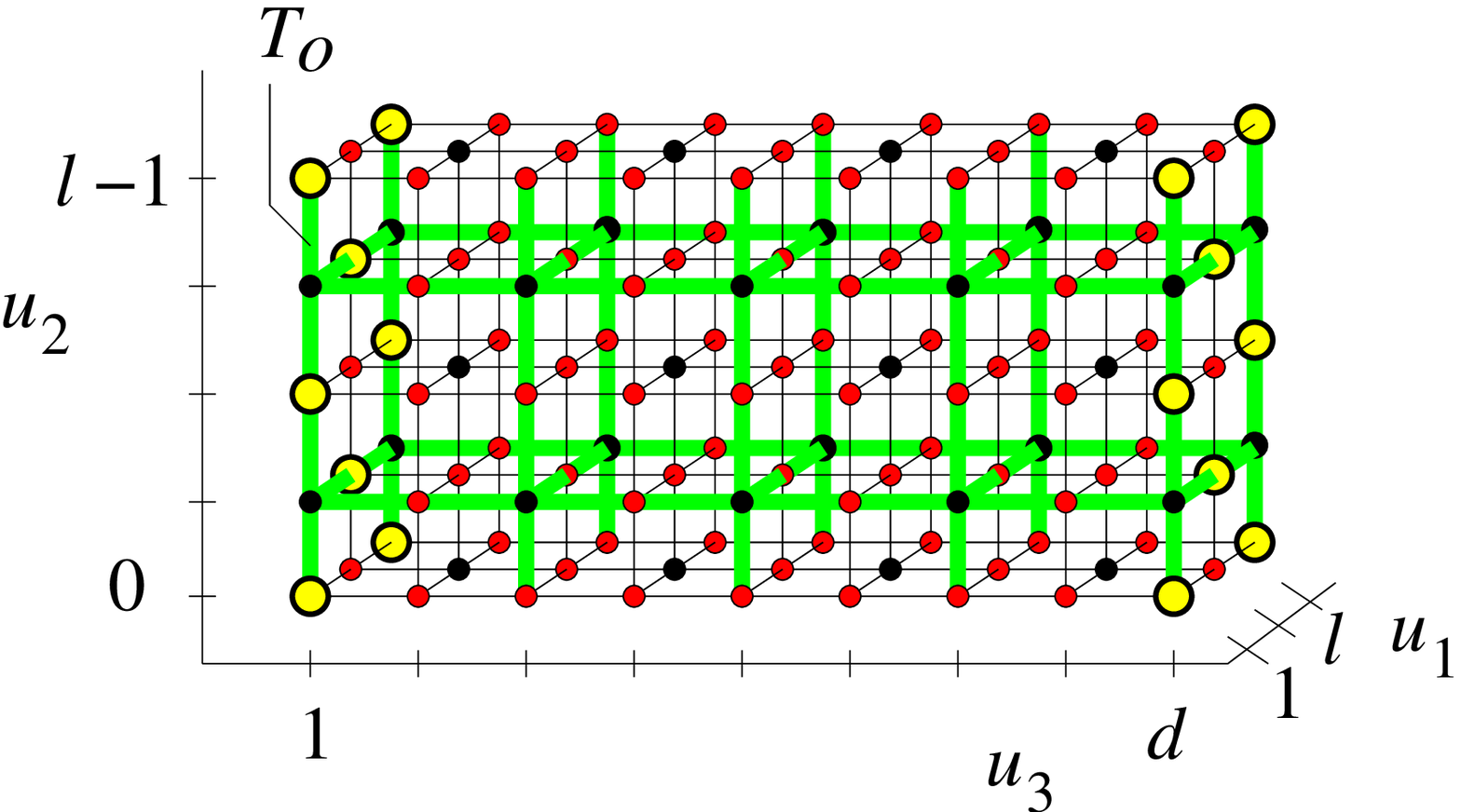, width=8.5cm}\\
    \epsfig{file=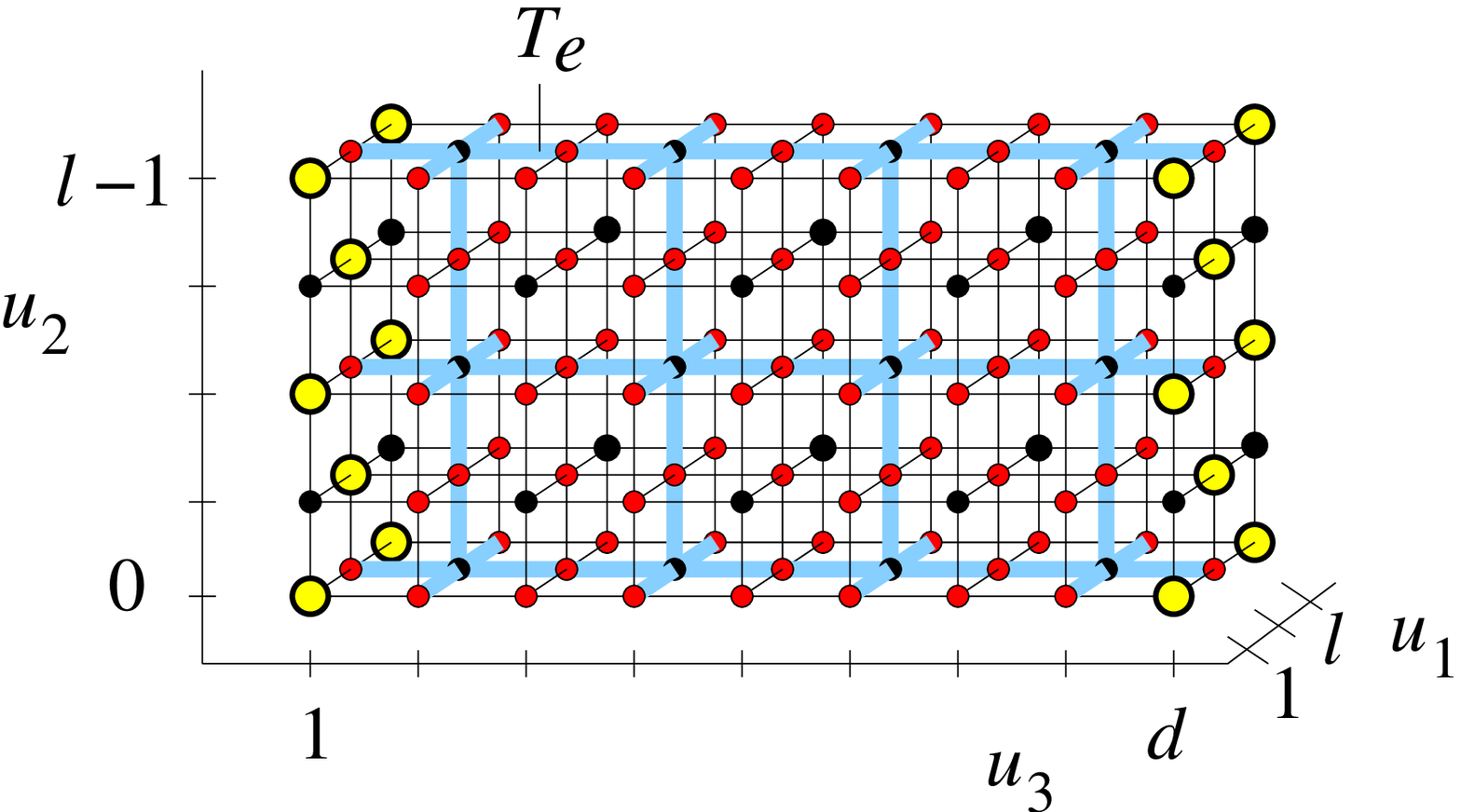, width=8.5cm}
    \caption{\label{measpatt}(color online) The measurement pattern on the 
      cluster 
  ${\cal{C}}$.  The sublattices $T_e$ and $T_o$ are displayed (thick lines). 
  For reference, the cluster lattice is also shown (thin lines) and
  the axis labeling shows the cluster coordinates.  
  Cluster qubits measured in  the $Z$-basis  (on the sites of $T_o$ and $T_e$) 
  are displayed in black, and  qubits measured in the $X$-basis (on the edges 
  of $T_o$ and $T_e$) are displayed in gray (red). The large circles to left 
  and to 
  the right denote the unmeasured qubits which form the encoded Bell pair. The 
  measurement pattern has a bcc symmetry.} 
  \end{center}
\end{figure}

Let us assume that the lengths $l$ and $d$ are odd~\footnote{There is no 
loss of generality here  since one can decrease the size of the lattice
by measuring all of the qubits on some of the 2D faces in the $Z$-basis.}. 
The Bell pair to be 
created between $A$ and $B$ will be encoded into subsets of qubits
\[
\ba{rcl}
L &=& \{ u=(o,e,1),(e,o,1) \in \calC\} \subset A,\\
R &=& \{ u=(o,e,d),(e,o,d) \in \calC\} \subset B.\\
\ea
\]
Each qubit $u\in \calC$ is measured either in the $Z$- or
 $X$-basis unless it belongs to $L$ or $R$. Denoting $M_X$ and $M_Z$ local 
$X$- and $Z$-measurements, we can now present the measurement pattern:
\begin{equation}
  \label{M1}
  \begin{array}{lr}
  M_Z: & \;\;\;\; \forall u \in V(T_e) \cup V(T_o),\\
  M_X: & \;\;\;\; \forall u \in E(T_e) \cup E(T_o)\backslash (L \cup R),\\
    \end{array}
\end{equation}
We denote the measurement outcome $\pm1$ at vertex $u$  by $z_u$ or
$x_u$, respectively.
A graphic illustration of the measurement patterns for the individual slices is
 given in Fig.~\ref{OneSlice}. 

Before we consider errors, let us discuss the effect of this
measurement pattern on a perfect cluster state. Consider some
fixed outcomes $\{x_u\}$, $\{z_u\}$ of local measurements and
let  $\psiLR$ be the reduced state of 
the unmeasured qubits $L$ and $R$. We will now show that $\psiLR$
is, modulo local unitaries, an encoded Bell pair, with each qubit encoded
by the planar code \cite{Kit2}, the planar
counterpart of the toric code \cite{Kit1}. The initial cluster state
obeys eigenvalue equations $K_u|\phi\rangle_{\cal{C}} =
|\phi\rangle_{\cal{C}}$. This implies for the
reduced state
\begin{equation}
  \label{plaquette}
  Z_{P,u} 
  \psiLR = 
  \lambda_{P,u} \psiLR,\;\;\forall \,\, u = (e,e,1),
\end{equation} 
where $Z_{P,u} = \bigotimes_{v \in \nb(u) \cap L}Z_v$ is a
plaquette ($z$-type)  stabilizer operator for the
planar code \cite{Kit2}.  The eigenvalue $\lambda_{P,u}$ 
depends upon the measurements outcomes as
$\lambda_{P,u}= x_u z_{(u_1,u_2,2)}$. 
Note that in the planar code the qubits live on
the edges of a lattice rather than on its vertices. The planar code
lattice is  distinct from the cluster lattice
${\cal{C}}$; see Figs.~\ref{OneSlice},\ref{measpatt}.

\begin{figure}[ht]
  \begin{center}
    \epsfig{file=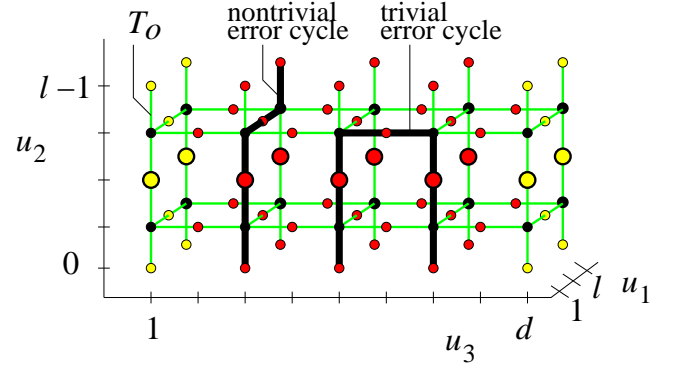, width=8.5cm}
    \caption{\label{cycl}(color online) A homologically nontrivial and a 
  homologically trivial error cycle on the lattice $T_o$. The nontrivial error 
  cycle stretches from one rough face to the opposite one while the trivial 
  error has both ends on the same face. Only the qubits belonging to $T_o$ are 
  shown and the qubits important for establishing the 
  $\overline{X}_A\overline{X}_B$-correlation are displayed enlarged.} 
  \end{center}
\end{figure}

From the equation  $\prod_{v \in \nb(u)} K_v\, |\phi\rangle_{\cal{C}}=
|\phi\rangle_{\cal{C}}$, for $u=(o,o,1)$, we obtain 
\begin{equation}
  \label{site}
  X_{S,u}\psiLR = \lambda_{S,u}
  \psiLR,\;\;\forall \,\, u = (o,o,1),
\end{equation}
where $X_{S,u}=\bigotimes_{v \in \nb(u) \cap L}X_v$ coincides with a site 
($x$-type) stabilizer operator for the planar code
\cite{Kit2}, and $\lambda_{S,u}= x_{(u_1,u_2,2)} z_u\prod_{v\in \nb_o(u)} z_v$,
where $\nb_o$ refers to a neighborhood relation on the sublattice $T_o$.
The code stabilizer operators in
Eq.~(\ref{plaquette}) and (\ref{site}) are algebraically
independent. There are $(l^2-1)/2$ code stabilizer generators for
$(l^2+1)/2$ unmeasured qubits, such that there exists one encoded
qubit on $L$. By direct analogy,  there is also one encoded qubit
located on $R$.

Next, we show that $\psiLR$ is an eigenstate of
$\ovl{X}_L \ovl{X}_R$ and $\ovl{Z}_L \ovl{Z}_R$, where $\ovl{X}$ and
$\ovl{Z}$ are the encoded Pauli operators $X$ and $Z$, respectively,
i.e. $\psiLR$ is an encoded Bell pair.  The encoded Pauli
operators \cite{Kit2} on $L$ and $R$ are $\ovl{X}_{L[R]} =
\bigotimes\limits_{u_1\,\, \text{odd}} X_{(u_1,u_2,1 [d])}$ for any
even $u_2$, and $\ovl{Z}_{L [R]} =
\bigotimes\limits_{u_2\,\,\text{even}} Z_{(u_1,u_2,1 [d])}$ for any
odd $u_1$.  To derive the Bell-correlations of $\psiLR$
let us introduce 2D slices
\[
\ba{rcl}
\Tsubsup{XX}{u_2} &=& \{ u = (o,u_2,o) \in \calC  \} \subset T_o,\\
\Tsubsup{ZZ}{u_1} &=& \{ u = (u_1,e,e) \in \calC  \} \subset T_e.\\
\ea
\]
The eigenvalue equation $\prod_{v \in
\Tsubsup{XX}{u_2}} K_v\, |\phi\rangle_{\cal{C}}= |\phi\rangle_{\cal{C}}$
with even  $u_2$ implies for the reduced state
\begin{equation}
  \label{XXcorr}
  \ovl{X}_L\ovl{X}_R\, \psiLR = \lambda_{XX}\,
  \psiLR,
\end{equation} 
with
$
  \lambda_{XX}=\prod_{v \in \Tsubsup{XX}{u_2+1} 
    \cup \Tsubsup{XX}{u_2-1} } z_v
  \prod_{v \in \Tsubsup{XX}{u_2}\backslash(L \cup R)} x_v.
$
Here and thereafter it is
understood that $x_u=z_u=1$ for all
$u \not \in {\cal{C}}$.
Similarly, from $|\phi\rangle_{\cal{C}}=\prod_{v \in \Tsubsup{ZZ}{u_1}}
 K_v\, |\phi\rangle_{\cal{C}}$, for $u_1$ odd, we obtain for the reduced state
\begin{equation}
  \label{ZZcorr}
  \ovl{Z}_L\ovl{Z}_R\, \psiLR = \lambda_{ZZ}\,
  \psiLR,
\end{equation} 
with $ \lambda_{ZZ}=\prod_{v \in \Tsubsup{ZZ}{u_1+1} \cup
\Tsubsup{ZZ}{u_1-1}} z_v \prod_{v \in \Tsubsup{ZZ}{u_1}} x_v$. Thus the
eigenvalue Eqs.~(\ref{plaquette}-\ref{ZZcorr}) show that the
measurement pattern of Eq.~(\ref{M1}) projects the initial perfect
cluster state into a state equivalent under local unitaries to the Bell pair, 
with each qubit encoded by the planar code. 

It is crucial that the measurement outcomes $\{z_u\}$ and $\{x_v\}$
are not completely independent. Indeed, for any vertex
$u\in T_o$ with $1<u_3<d$  the eigenvalue equation
$\prod_{v \in \nb(u)} K_v\, |\phi\rangle_{\cal{C}} = |\phi\rangle_{\calC}$
implies the constraint
\begin{equation}
  \label{Sy1}
\prod_{v \in  \nb(u)} x_v \cdot \prod_{w\in \nb_o(u)} z_w =1 .
\end{equation}
Analogously, for any vertex $u\in T_e$ one has a constraint
\begin{equation}
  \label{Sy2}
\prod_{v \in  \nb(u)} x_v \cdot \prod_{w\in \nb_e(u)} z_w = 1,
\end{equation}
where $\nb_e$ refers to a neighborhood relation on the lattice $T_e$.
Thus there exists one syndrome bit for each vertex of $T_e$ and $T_o$,
(with exception for the vertices of $T_o$ with $u_3=1$ or $u_3=d$).

What are the errors detected by these syndrome bits?
Since we have only $Z$-errors (for generalization, see remark 1), 
only the $X$-measurements are affected by
them. Each $X$-measured qubit is either on an edge of $T_o$ or
$T_e$.  Thus, we can identify the locations of the elementary errors
with $E(T_o)$ and $E(T_e)$. From the equations Eq.~(\ref{Sy1},\ref{Sy2}),
each error located on an edge creates a syndrome at its end vertices.

Let us briefly compare with \cite{TCDS}. Therein, independent local $X$-and $Z$-errors were considered for storage whose correction runs completely independently. The $X$-errors in this model correspond to our $Z$-errors on qubits in $E(T_e)$, and the $Z$-storage errors to our $Z$-errors on qubits in $E(T_o)$, if the $X-$ and $Z-$error correction phases in \cite{TCDS} are pictured as alternating in time.

The syndrome information provided by Eqs.~(\ref{Sy1},\ref{Sy2}) is not
yet complete.  There are two important issues to be
addressed: (i) There are no syndrome bits at the vertices of $T_o$ with
$u_3=1$ or $u_3=d$; (ii) Edges of $T_e$ with $u_3=1$ or $u_3=d$ have only
one end vertex, so errors that occur on these edges create only one
syndrome bit.  Concerning (i), to get the missing syndrome bits we will measure 
eigenvalues $\lambda_{P,u}$ and $\lambda_{S,u}$ for 
the plaquette and the site stabilizer
operators living on the faces $A$ and $B$,
see Eqs.~(\ref{plaquette},\ref{site}).
Such measurements are local operations within $A$ or within $B$, so they
can not increase entanglement between $A$ and $B$.
For any $u=(o,o,1)$ or $u=(o,o,d)$ it follows from Eq.~(\ref{site}) that
\be
\label{Sy3}
\begin{array}{rclr}
  \ds{\lambda_{S,u}\, x_{(u_1,u_2,2)}\,  z_u \prod_{v\in \nb_o(u)}z_v} &=& 1,& 
  \mbox{for } u_3=1,\\ 
  \ds{\lambda_{S,u}\,x_{(u_1,u_2,d-1)}\, z_u \prod_{v\in \nb_o(u)}z_v} &=& 1, &
  \mbox{for } u_3=d. 
\end{array}
\ee
For any vertex $u=(o,o,1)$ or $u=(o,o,d)$ there are several 
edges of the lattice $T_o$ incident to $u$. 
It is easy to see that a single $Z$-error that occurs on any of these edges
changes a sign in Eqs.~(\ref{Sy3}). 
Thus, these two constraints yield the syndrome bits living at the vertices 
$u=(o,o,1)$ and $u=(o,o,d)$, so the issue~(i) is addressed.
Concerning (ii), we make use of Eq.~(\ref{plaquette}) and obtain
\be
\label{Sy5}
\begin{array}{rclr}
  \lambda_{P,u}\,  x_u z_{(u_1,u_2,2)} &=& 1,& \mbox{for any }
  u=(e,e,1),\vspace{2mm} \\
  \lambda_{P,u}\,  x_u z_{(u_1,u_2,d-1)} &=& 1,& \mbox{for any } 
  u=(e,e,d).
\end{array}
\ee
Since we have only $Z$-errors, the eigenvalues $\lambda_{P,u}$ and
the outcomes $z_{(u_1,u_2,2)}$, $z_{(u_1,u_2,d-1)}$ are not affected by
errors. Thus the syndrome bits Eqs.~(\ref{Sy5}) are equal to $-1$
iff an error has occurred on the edge $u=(e,e,1)$ or $u=(e,e,d)$ of the
lattice $T_e$. Since each of these errors shows itself in a corresponding 
syndrome bit which is not affected by any other error, we can reliably 
identify 
these errors.  This is equivalent to actively correcting them with unit 
success probability. We can therefore assume in the subsequent analysis 
that no errors occur on the edges $(e,e,1)$ and $(e,e,d)$, which concludes 
the discussion of the issue (ii).

As in \cite{Kit1}, we
define an error chain ${\cal{E}}$ as a collection of edges where an 
elementary error has occurred. Each of the two lattices $T_e$ and $T_o$ has its own error chain. An error chain ${\cal{E}}$ shows a syndrome only at its
boundary $\partial({\cal{E}})$, and errors with the same boundary thus
have the same syndrome. One may identify an error ${\cal{E}}$ only
modulo a cycle $D$, ${\cal{E}}^\prime = {\cal{E}} + D$, with
$\partial(D)=0$.

There are homologically trivial and nontrivial cycles. A cycle $D$ is
trivial if it is a closed loop in $T_{o}$ ($T_e$), and homologically
nontrivial if it stretches from one rough face in $T_{o}$ ($T_{e}$) to
another. A rough face here is the 2D analogue of a rough edge on a
planar code \cite{Kit2}. The rough faces of $T_o$
are on the upper and lower side of ${\cal{C}}$, and the rough faces of
$T_e$ are on the front and back of ${\cal{C}}$ (recall that no errors
occur on the left and right rough faces of $T_e$).

Let us now study the effect of error cycles on the identification of
the state $\psiLR$ from the measurement outcomes. We only discuss
the error chains on $T_o$ here, which potentially affect the
eigenvalue Eq.~(\ref{XXcorr}). The discussion of the error chains in
$T_e$---which disturb the $\ovl{Z}_L\ovl{Z}_R$-correlations---is
analogous. An individual qubit error on $v\in {\cal{C}}$ will modify
the $\ovl{X}_L\ovl{X}_R$ correlation of $\psiLR$ if it either
affects $\ovl{X}_L$, $\ovl{X}_R$ or $\lambda_{XX}$. That happens if $v
\in T_{XX}^{(u_2)}$. Now, the vertices in $T_{XX}^{(u_2)}$ correspond
to edges in $T_o$. If an error cycle $D$ in $T_o$ is homologically
trivial, it intersects $T_{XX}^{(u_2)}$ in an even number of
vertices; see Fig.~\ref{cycl}. This has no effect on the eigenvalue
Eq.~(\ref{XXcorr}). However, if the cycle is homologically nontrivial,
i.e. if it stretches between the upper and lower face of ${\cal{C}}$,
then it intersects $T_{XX}^{(u_2)}$ in an odd number of vertices. This
does modify the eigenvalue Eq.~(\ref{XXcorr}) by a sign factor of
$(-1)$ on the l.h.s., which leads to a logical error.
Therefore, for large system size, we require the probability 
of misinterpreting the syndrome by a nontrivial cycle to be negligible 
\cite{TCDS}:
\begin{equation}
  \sum_{\cal{E}} \text{prob}({\cal{E}}) \sum_{D\; \text{nontrivial}}
  \text{prob}({\cal{E}}+D|\,{\cal{E}})\, \approx \,0.
\end{equation}
We have now traced back the problem of reconstructing an encoded Bell pair
$\psiLR$ to the same setting that was found in \cite{TCDS}
to describe fault-tolerant data storage with the toric code. Via the
measurement pattern Eq.~(\ref{M1}), we may introduce two lattices
$T_o$, $T_e$ such that 1) Syndrome bits are located on the vertices of
these lattices, 2) Independent errors live on the edges and show a
syndrome on their boundary, 3) Only the homologically nontrivial
cycles give rise to a logical error.  This error model can be 
mapped onto a random plaquette $Z_2$-gauge field theory in 3 dimensions 
\cite{TCDS,WHP} which undergoes a phase transition between an
ordered low temperature and a disordered high temperature phase. In the limit of $l,d \longrightarrow \infty$, full error-correction is possible in the low temperature phase. 

In our setting, the error probabilities for all edges are equal to
$p$. For this case the critical error probability has been computed numerically in a lattice simulation \cite{PT_Nishimori}, $p_c=0.033\pm 0.001$. This value  corresponds, via Eq.~(\ref{Eprob}), to $T_c=(0.296\pm 0.003)\Delta$.

{\em{Remarks:}} 1) The error model equivalent to Eq. (\ref{CS}), i.e. 
$Z$-errors only, is very restricted. We have a physical motivation for this model, but we would like to point out that the very strong assumptions we have made about the noise are not crucial to our result of the threshold error rate being non-zero. One may, for example, generalize the error model from a dephasing channel to a depolarizing channel, with $p_x=p_y=p_z=p^\prime/3$.
Then, two changes need to be addressed, those in the bulk and those on the faces $L$ and $R$. Concerning the faces, the errors on the rough faces to the left and right of $T_e$ can no longer be unambiguously identified by measurements of the code stabilizer (\ref{Sy5}), which raises the question of whether---for depolarizing errors---it may be these surface errors that set the threshold for long-range entanglement. This is not the case. To see this, note that two slices of 2D cluster states may be attached to the left and right of ${\cal{C}}$, at $u_3=0,-1$ and $u_3=d+1,d+2$. The required operations are assumed to be perfect. They do not change the localizable entanglement between the left and right side of the cluster ${\cal{C}}$ because they act locally on the slices $-1 .. 1$ and $d .. d+2$, respectively. The subsets $A$ and $B$ of spins are re-located to the slices $-1$ and $d+2$, with the corresponding changes in the measurement pattern. The effect of this procedure is that the leftmost and rightmost slice of the enlarged cluster are error-free \footnote{ The following operations are required to attach a slice: (I) $\Lambda(Z)$-gates within the slice, (II) $\Lambda(Z)$-gates between the slice and its next neighboring slice, (III) $X$- and $Z$-measurements within the slice, see Fig.~1. All these operations are assumed to be perfect, and the errors on slices $1$ and $d$ are not propagated to slices $-1$ and $d+2$ by the $\Lambda(Z)$-gates (II).}, and only the bulk errors matter. 

\begin{figure}
  \begin{center}
    \epsfig{file=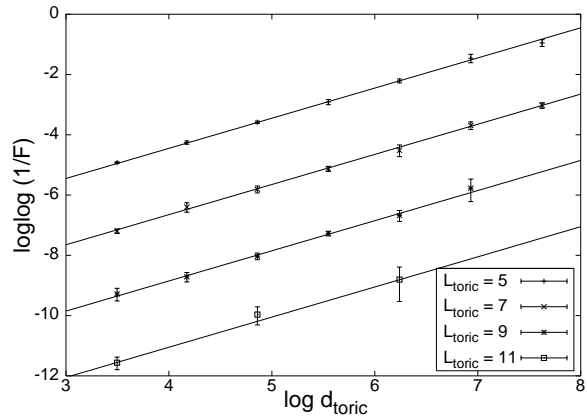,width=8cm} 
    \caption{\label{phaseErrorsFigure}This figure plots data for simulations 
      of error correction on an $L
      \times L \times d_\text{toric}$ lattice, with periodic boundary
      conditions in the first two directions, for various $L$
      and $d_\text{toric}$ ($d=2d_\text{toric}+1$, $l=2L$). 
      The error rate is $p=0.01$. The logs are
      base $e$.  Two standard deviations
      above and below the computed values (as given by statistical noise due
      to the sample sizes) are shown by the error bars.  The solid lines
      each have slope one, and they are spaced equally apart.  This lends
      good support to the model of fidelity $F \sim \exp( -d k_1 \exp( -l
      k_2))$ for error rates below threshold.}
  \end{center}
\end{figure}

Concerning the bulk, note that the cluster qubits  measured in $Z$-basis 
serve no purpose and may be left out from
the beginning. Then, the considered lattice for the initial cluster
state has a bcc symmetry and double spacing. The lattices $T_o$, $T_e$
remain unchanged. Further, $X$-errors are absorbed in the $X$-measurements 
and $Y$-errors act like $Z$-errors, such that we still map to the original 
$\mathbb{Z}_2$ gauge model \cite{TCDS} at the Nishimori line. The threshold 
for local depolarizing 
channels applied to this configuration is thus $p^{\prime}_c=3/2 \,p_c=
4.9\%$. In addition, numerical simulations performed for
the initial simple cubic cluster and depolarizing channel  
yield an estimate of the critical error probability of 
$p^{\prime\prime}_c=1.4\%$. \\ 2)
Finite size effects.  We carried out numerical simulations of error
correction on an $l \times l \times d$ lattice with periodic boundary
conditions (as opposed to the open boundary conditions of the planar
codes within the cluster state).  For differing error rates below the
threshold value of $2.9\%$ \cite{WHP}, we found good
agreement for the fidelity $F$ between the perfect and the error-corrected encoded Bell state with the model $F \sim \exp( -d k_1 \exp( -l
k_2))$.  Some data is shown in Fig. \ref{phaseErrorsFigure}
corresponding to a $Z$-error rate of $1.0\%$.  Provided that planar
codes and toric codes have similar behavior away from threshold, our
simulations suggest that, in order to achieve constant fidelity, the
length $l$ specifying the surface code need only scale
\emph{logarithmically} with the distance $d$.\\ 3)  For even $d$, 
the construction presented above 
can be used to mediate an encoded conditional $Z$-gate on 
distant encoded qubits located on slices $1$ and $d$.

 \section{Upper bound}
\label{sec:UB}

In this section we analyze the high-temperature behavior of thermal cluster
states and find an upper bound on the critical temperature $T_c$. 
Our analysis is based on the isomorphism between cluster states and the
so-called Valence Bond Solids (VBS) pointed out by Verstraete and Cirac
in~\cite{Verstraete_Cirac_VBS} which can easily be generalized to a finite
temperature. 
\begin{figure}[ht]
\begin{center}
  \epsfig{file=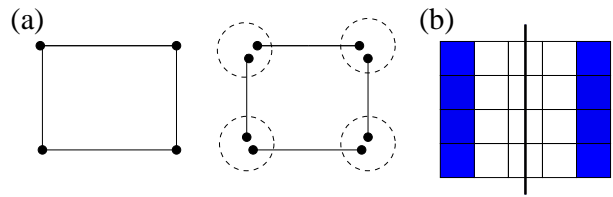, width=8cm}
  \caption{\label{VBpict}(color online) (a) Correspondence between physical 
and virtual qubits. Domains are shown by dashed lines.
(b) A bipartite cut of a cubic lattice. The regions $A$ and $B$ are
highlighted.}
\end{center}
\label{fig:VBS}
\end{figure}

With each physical qubit $u\in \calC$ we associate 
a domain $u.*$ of $d(u)$ virtual qubits,
where $d(u)=|\nb(u)|$ is the number of nearest neighbors of $u$ 
(see Fig.~\ref{VBpict}~(a)).
Let us label virtual qubits from a domain $u.*$ as $u.v$, $v\in \nb(u)$.
Denote $E$ to be the set of edges of the lattice $\calC$ and define a thermal 
VBS state $\rhosub{V\!BS}$ as
\begin{equation}\label{VBS}
\rhosub{V\!BS}= \prod_{e=(u,v)\in E}
\frac14 \left( I + \omega_e X_{u.v}Z_{v.u}\right)
\left( I + \omega_e Z_{u.v}X_{v.u}\right).
\end{equation}
Here $\{\omega_e\}$ are arbitrary weights such that $0\le \omega_e \le 1$.
It should be emphasized that $\rhosub{V\!BS}$ is a state of virtual qubits
rather than physical ones. 
Our goal is to convert $\rhosub{V\!BS}$ into $\rhosub{CS}$
by local transformations mapping a domain $u.*$ into a single qubit $u\in \calC$.
The following theorem is a straightforward generalization of the
Verstraete and Cirac construction (here we put $\Delta/2=1$).
\begin{theorem}\label{theorem1}
Let $\rhosub{CS}$ be a thermal cluster state on the 3D cubic lattice $\calC$
at a temperature $T\equiv \beta^{-1}$. 
Consider a thermal VBS state $\rhosub{V\!BS}$
as in Eq.~(\ref{VBS}) such that the weights $\omega_e$ satisfy
\begin{equation}\label{div}
\prod_{v\in \nb{(u)}}
\omega_{(u,v)}\, \ge \tanh{(\beta)} \quad \mbox{for each} \quad u\in \calC.
\end{equation}
Then $\rhosub{V\!BS}$ can be converted into $\rhosub{CS}$ by
applying a completely positive transformation
${\cal{W}}_u$ to each domain $u.*$,
\begin{equation}\label{CS&VBS}
\rhosub{CS} =  {\cal{W}}( \rhosub{V\!BS} ), \quad  {\cal{W}}=\bigotimes_{u\in \calC} {\cal{W}}_u.
\end{equation}
\end{theorem}
Let us first discuss the consequences of this theorem. Note that each 
edge $e\in E$ of $\rhosub{V\!BS}$ carries a two-qubit state  
\begin{equation}\label{bond state}
\rho_e =\frac14 (I + \omega_e X_1 Z_2)(I + \omega_e Z_1 X_2).
\end{equation}
The Peres-Horodecki partial transpose criterion~\cite{Peres_PPT,Horodecki_PPT}
tells us that $\rho_e$ is separable if and only if $\omega_e \le \sqrt{2}-1$. 
Consider a bipartite cut of the lattice by a hyperplane of
codimension 1 (see Fig.~\ref{VBpict}~(b)). 
We can satisfy
Eq.~(\ref{div}) by setting $\omega_e=\tanh{(\beta)}$ for all edges crossing
the cut and setting $\omega_e=1$ for all other edges.
Clearly, the state $\rhosub{V\!BS}$ is bi-separable  whenever
$\tanh{(\beta)}\le \sqrt{2}-1$. But bi-separability of $\rhosub{V\!BS}$
implies bi-separability of $\rhosub{CS}$. We conclude that the localizable
entanglement between the regions $A$ and $B$ is zero whenever $\tanh{(\beta)}\le \sqrt{2}-1$,
which yields the upper bound on $T_c$ presented earlier. 

{\it Remarks:} We can also satisfy Eq.~(\ref{div}) by setting $\omega_e=\omega$
for all $e\in E$, with $\omega^6 = \tanh{(\beta)}$. 
This choice demonstrates  that
$\rhosub{CS}$ is completely separable for $\tanh{(\beta)}<(\sqrt{2}-1)^6$
(that is $T\approx 200$).
It reproduces the upper bound~\cite{MacE}  of D\"ur and Briegel
on the separability threshold error rate for cluster states.

In the remainder of this section we prove Theorem~\ref{theorem1}.
Consider an algebra $\calA_u$ of operators acting on some particular
domain $u.*$. It is generated by the Pauli operators
$Z_{u.v}$ and $X_{u.v}$ with $v\in \nb{(u)}$. 
The transformation ${\cal{W}}_u$ maps
$\calA_u$ into the one-qubit algebra generated
by the Pauli operators $Z_u$ and $X_u$. First, we choose
\[
{\cal{W}}_u(\eta)= W_u^\dag\, \eta \, W_u, \quad 
W_u = |0^{\otimes d(u)}\ra\la 0| + |1^{\otimes d(u)}\ra\la 1|.
\]
One can easily check that
\begin{equation}\label{conj2}
W_u^\dag Z_{u.v} = Z_u W_u^\dag \quad \mbox{and} \quad
Z_{u.v} W_u = W_u Z_u,
\end{equation}
for any $v\in \nb{(u)}$. As for commutation relations between
$W_u$ and $X_{u.v}$ one has
\begin{equation}\label{conj3}
  \begin{array}{rcl}
    W_u^\dag \, \left( \prod_{v\in \nb{(u)}} X_{u.v}\right) 
      \, W_u &=& X_u,\\
    W_u^\dag \, \left( \prod_{v\in S} X_{u.v}\right) \, W_u 
    &=&0,
  \end{array}
\end{equation}
for any non-empty proper subset $S\subset \nb{(u)}$. 
Taking ${\cal{W}}=\bigotimes_{u\in \calC} {\cal{W}}_u$ and using 
Eqs.~(\ref{conj2}), (\ref{conj3}) one can easily get
\begin{equation}\label{quasy_CS}
{\cal{W}}(\rhosub{V\!BS}) = \frac{1}{4^{|E|}} \prod_{u\in \calC}
\left( I + \eta_u K_u \right), \quad
\eta_u = \prod_{v\in \nb{(u)}} \omega_{(u,v)}.
\end{equation}
We can regard the state in Eq.~(\ref{quasy_CS})
as a thermal cluster state with a local temperature
$\tanh{(\beta_u)}\equiv \eta_u$ depending upon $u$. 
The inequality of Eq.~(\ref{div}) implies that $\beta_u\ge \beta$ for all $u$.
To achieve a uniform temperature distribution $\beta_u=\beta$ one can 
intentionally apply local $Z$-errors with properly chosen
probabilities.

\section{Conclusion}

Thermal cluster states in three dimensions exhibit a transition from
infinite to finite entanglement length at a non-zero transition
temperature $T_c$.  We have given a lower and an upper bound to $T_c$,
$0.3\,\Delta\leq T_c\leq 1.15\,\Delta$ ($\Delta=$ energy gap of the
Hamiltonian). The reason for $T_c$ being non-zero is an intrinsic
error-correction capability of 3D cluster states. We have devised an
explicit measurement pattern that establishes a connection between
cluster states and surface codes. Using this, we have described how to
create a Bell state of far separated encoded qubits in the
low-temperature regime $T<0.3\, \Delta$, making the entanglement
contained in the initial thermal state accessible for quantum 
communication and computation.

\begin{acknowledgments}
We would like to thank Hans Briegel and Frank Verstraete for bringing to our
attention the problem of entanglement localization in thermal
cluster states.
This work was supported by the National Science Foundation
under grant number EIA-0086038.
\end{acknowledgments}



\begin{thebibliography}{99}
\bibitem{Bell}
J.S. Bell, Physics {\bf{1}}, 195 (1964).

\bibitem{Aspect}
A. Aspect, P. Grangier and G. Roger, Phys. Rev. Lett. {\bf{47}}, 460 (1981).

\bibitem{Schr}
E. Schr{\"o}dinger, Naturwissenschaften {\bf{23}}, 807-812, 823-828, 844-849 (1935).

\bibitem{Rep}
H.J. Briegel, W. D{\"u}r, J.I. Cirac, and P.Zoller, Phys. Rev. Lett. {\bf{81}}, 5932 (1998).


\bibitem{LE}
F. Verstraete, M. Popp and J.I. Cirac, Phys. Rev. Lett. {\bf{92}}, 027901 (2004).

           

\bibitem{Aha}
D. Aharonov, quant-ph/9910081 (1999).

\bibitem{Verstraete_Delgado_Cirac_04}  F. Verstraete, M.A. Mart\'in-Delgado, J.I. Cirac,
Phys. Rev. Lett. {\bf 92}, 087201 (2004).

\bibitem{AKLT} I. Affleck, T. Kennedy, E.H.~Lieb, and H.~Tasaki,
Commun. Math. Phys. {\bf 115}, 477 (1998).

\bibitem{PP}
J.K. Pachos and M.B. Plenio, Phys. Rev. Lett. {\bf{93}}, 056402 (2004).

\bibitem{Kay}
A. Kay {\em{et al.}}, quant-ph/0407121 (2004).

\bibitem{BR}
H.J. Briegel and R. Raussendorf, Phys. Rev. Lett. {\bf{86}}, 910 (2001).

\bibitem{TCDS}
E. Dennis, A. Kitaev, A. Landahl, and J. Preskill, quant-ph/0110143 (2001).

\bibitem{Kit2}
S. Bravyi, A. Kitaev, quant-ph/9811052 (1998).

\bibitem{Kit1}
A. Kitaev, quant-ph/9707021 (1997).

\bibitem{WHP}
C. Wang, J. Harrington and J. Preskill, Annals Phys. {\bf{303}}, 31 (2003), quant-ph/0207088 (2002).

\bibitem{PT_Nishimori}
T. Ohno, G. Arakawa, I. Ichinose and T. Matsui, quant-ph/0401101 (2004).

\bibitem{Verstraete_Cirac_VBS} 
F. Verstraete and J.I. Cirac,
quant-ph/0311130 (2003).

\bibitem{Peres_PPT}
A. Peres, Phys. Rev. Lett.  {\bf 77}, 1413 (1996).

\bibitem{Horodecki_PPT}
M. Horodecki, P. Horodecki, and R. Horodecki,\\ Phys.~Lett.~A
{\bf 223}, 1 (1996).

\bibitem{MacE}
W. D{\"u}r and H.J. Briegel, Phys. Rev. Lett. {\bf{92}}, 180403 (2004). 

\end{thebibliography}
\end{document}